# A conceptual framework of Intelligent Management Control System for Higher Education

**Helena Dudycz[1], Marcin Hernes[1], Zdzieław Kes[1], Eunika Mercier-Laurent[2], Bartłomiej Nita[1], Krzysztof Nowosielski[1], Piotr Oleksyk[1], Mieczysław L. Owoc[1], Rafał Palak[3], Maciej Pondel[1], Krystian Wojtkiewicz[3]**

[1]Wrocław University of Economics and Bussiness, Komandorska 118/120, 53-345 Wrocław, Poland

{helena.dudycz, marcin.hernes, zdzisław kes, bartłomiej.nita, krzysztof.nowosielski, piotr.oleksyk, mieczysław.owoc, maciej.pondel}@ue.wroc.pl

[2] University of Reims Champagne Ardennes, France

eunika@innovation3d.fr

[3] Wroclaw University of Science and Technology, Poland

{rafal.palak, krystian.wojtkiewicz}@pwr.edu.pl

**Abstract.** The utilization of management control systems in university management poses a considerable challenge because university's strategic goals are not identical to those applied in profit-oriented management. A university's management control system should take into account the processing of management information for management purposes, allowing for the relationships between different groups of stakeholders. The specificity of the university operation assumes conducting long-term scientific research and educational programmes. Therefore, the controlling approach to university management should considerat long-term performance measurement as well as management in key areas such as research, provision of education to students, and interaction with the tertiary institution's socio-economic environment.This paper aims to develop a conceptual framework of the Intelligent Management Control System for Higher Education (IMCSHE) based on cognitive agents. The main findings are related to developing the assumption, model, and technological basis including the artificial intelligence method.

## 1. Introduction

The use of management control systems (MCS) in university management is a significant challenge because the universities' strategic objectives are not the same as those of profit-oriented businesses. Moreover, unlike enterprises, which mainly attach importance to increasing value for the owners, the competitive advantage - and thus the position - of a university among institutions which focus on scientific and educational activity stems from other sources and is oriented towards meeting the needs of a wide range of stakeholders. It is crucial to translate a university's strategy into its operation, which is intended to serve the purpose of enabling all members of the academic community to understand the strategy while streamlining the process of its implementation in the course of the tertiary institution's day-to-day activity. A university's controlling system structure is closely related to its strategic goals and assumed directions of development. The triple helix concept, promoted by H. Etzkowitz and L. Leydesdorff [1] can be used as a basis for defining the mission and strategy of a university. According to them, the triple helix is a model of relations occurring in the process of the creation of knowledge between three types of entities: scientific centres (academia), the industrial environment (industry), and public administration (state). The knowledge-based economy requires creative cooperation of modern universities with public administration and the economic environment. The so-called quadruple helix and even quintuple helix concepts are being postulated today [2]. The triple helix model is enriched by the media, civil society, and the natural environment.

The MCS of a university should allow for the processing of management information for the executive bodies' needs, taking into account the relations between different stakeholder groups. The specificity of a university's operation assumes conducting long-term research and offering long-term educational programmes. Therefore, university management's controlling approach should allow for long-term performance measurement and management in critical areas, such as conducting research, educating students, and cooperating with the socio-economic environment.

The design of MCS to support university management should include internal and external information. Processing internal information is complicated due to the multidimensionality of information aggregation objects and the wide range of stakeholders. Access to prospective information allowing to assess the expected changes in operating conditions and determine future directions of action is limited because scientific research may produce uncertain results (i.e. there are no such methods). A useful tool for any university is to create scenarios for future decisions that can be generated to ensure the adopted development strategy's implementation. Moreover, as far as the decision-making process is concerned, it is necessary to process external information concerning factors including the following:

- the demographic situation,
- changes in legal regulations,

- the demand for graduates with new qualifications,
- competitive entities' activity,
- changes in the methods of education.

Managing a university is a process of making decisions whose accuracy determines the effectiveness and efficiency of its educational and research activities. Therefore, analyses carried out under the MCS should enable the following, among other things:

- evaluation of the university's ability to meet the requirements of the regulations governing higher education,
- identification of the financial needs sources of research and sources of its funding,
- symptoms of threat and the risk of a change in the competitive position,
- directions of changes in the side areas of operation.

A carefully developed and detail-oriented internal information system, constituting a response to signals coming from the environment, plays a vital role in supporting university management. It is necessary to apply a detailed assessment of all aspects of the university's operation. The most common method of managing performance in private and public sectors, including universities, is the Balanced Scorecard. The first generation of the Balanced Scorecard [3] was treated exclusively as a method of measuring performance, while nowadays, it is a much more advanced concept, one that is continuously being developed. Today, R.S. Kaplan and D. P. Norton argue that their proposal can provide a framework for performance management geared towards implementing a strategy using intelligent control systems.

The functioning of the MCS involves the processing of both structured and unstructured knowledge. The artificial intelligence methods and tools can be used in this purpose.

This paper aims to develop a conceptual framework of an Intelligent Management Control System for Higher Education (IMCSHE) based on cognitive agents.

The main contribution is related to the following:

- The application of cognitive agents in a Management Control System for Higher Education,
- A proposal for extending the Learning Intelligent Distribution Agent Architecture with a machine learning module.

## 2. Related works

During the theoretical foundations' analysis, attention was paid to the issues raised in studies dedicated to Management Control Systems (MCS) solutions implemented at universities, emphasizing the applications being within this study's scope. The overwhelming majority of the studies concerns the characteristics and evaluation of MSC systems used. Baid [4] researched types of control in schools. Its findings show a positive relationship between teacher control, student control and the institution's performance. Al.-Tarawneh [5] conducted a study involving university professors and students in the context of MSC use. In his view, professors who have less experience in higher education tend to use MSC more than those with many years of experience [5]. Beuren and Teixeira [6] researched the structure and functions of MSC in assessing performance at private universities. Based on the results of the questionnaires sent to managers responsible for the strategy of educational institutions, it was found that the overall evaluation result for MCS reached an average of 3.62 on a 5-degree scale. According to Tanveer and Karim [7], universities must implement performance management (PM) procedures to improve their performance and align their individual goals with the university's strategy. The evaluation must consider the achievements of the researchers and the administration to an equal extent.

A much smaller number of articles included information on the IT solutions being applied. The authors Cantele, Martini, and Campedelli [8] showed in their research an insignificant relationship between the degree of development of MCS and some specific features of state-run universities, such as size, complexity, degree of expenditure rigidity, and availability of human resources assigned to administrative activities. Besides, they found that the planning and control tools used at universities include modules that enable the following:

- economic and financial simulations, including planning under different scenarios;
- processing information on the university's strategy, mission, and objectives;
- building long-term business plans that define strategies, goals, programmes, initiatives, and projects, as well as their implementation schedules, project managers, allocated financial and human resources, and objectives to be achieved;
- using management dashboards which provide performance indicators that can be used for both internal management and external communication;
- cost calculation in sections such as master's studies, research projects, activities and processes, responsibility centres;
- budgeting and operational planning;
- deviation analysis.

Labas and Bacs [9] present an MCS concept that includes a software structure. The MCS should be at the junction of field-specific systems (e.g. fixed assets management, financial management, financial accounting, or human resources management) and the management information system (MIS). Based on preliminary research carried out, it can be concluded that there is a cognitive gap in IT support for MCS in organizations involved in academic teaching.

Z Alach [10] proposes a model for assessing the maturity of performance management systems in higher education controlling, which includes seven elements. This model was built based on normative research conducted by many authors. Z Alach integrated these recommendations and empirically verified them, grouping all criteria into three sets. The first group concerns the use of information on performance in university management and on whether the obtained results stem from the use of this information. The second group is focused on designing a performance management system and its internal coherence. The last group refers to the elements that determine the system's usefulness. McCormack et al. [11] propose using 17 measures for the evaluation and design of a university's controlling system. All the measures are divided into four categories: operational activity, monitoring, targeting of measures, and motivation systems.

## 3. Research methodology

In our research, we use the Design Science Research Process (DSRP) methodology [12]. This methodology involves the realization of research works as process-dependent steps responsible for implementing basic tasks and delivering primary work artefacts. This methodology is based on five necessary steps. The first, "Awareness of problem", focuses on preparing a preliminary proposal for a new research topic. It results from the analysis of already available achievements and results in related scientific disciplines and industry (in our paper this phase is realized in the introduction. The second, development of the proposal ("Suggestion"), follows the initial proposal and is a creative step to define new models, methods, and functionalities based on new configurations of existing or new elements. This phase is realized in the related works section. The third, constituting further development of the proposal ("Development"), is where the initial research is developed and extended. The development is performed in the Framework of the system section. The fourth, "Evaluation", is when the prepared artefacts are evaluated according to the adopted criteria. The fifth, "Conclusion", can be the end of a given research cycle or end the entire research undertaking. The evaluation and conclusion are related in the Conclusions section. As outlined, each stage ends with a specific result, which then forms the basis for and enables the next step in the process. Besides, this methodology assumes that it is possible to return to the previous steps under the observed restrictions. We use the literature of subject analysis, as a research method in the Awareness of the problem and the Suggestion stages. In the development stage, we use such research methods, as modelling, design and implementation and observation of phenomena's. The Evaluation we use research experiment method. Conclusion phase we based on synthesis, analysis, induction and deduction research methods.

## 4. The framework of the system

The Intelligent Management Control System for Higher Education (IMCSHE) aims to provide information for the UEW's Balanced Scorecard (BSC). The scorecard includes four perspectives focusing on the following information:

### *The financial perspective (F):*

- financial plans including a material and financial plan, the central budget of the university, partial budgets (budgets of subunits, tasks, or projects);
- control aspects in terms of income and expenses (in cross-section of subunits, tasks and projects) and financial and economic indicators.

### The customer perspective (C):

- student segmentation;
- change dynamics in student segments (degrees, types of studies, fields of study);
- quality of education from the perspective of students and graduates (tracking the path of graduates' professional development) broken down into fields of study, specialties, modules, and subjects;
- education quality indicators.

### The internal processes perspective (P):

- human resources potential, including the structure and dynamics of employment broken down into degrees and academic titles, positions held, and functions performed;
- scientific potential, including the number and quality of publications (IF and quality of publications), citation indicators, experience in research projects, the number of promoted doctoral students, the level of scientific research commercialization, the impact of scientific activity on the functioning of the society and economy, research infrastructure, as well as the organization of conferences and scientific seminars;
- educational potential, including flexibility in adopting the educational offer to market needs, employee evaluation from the perspective of both, the student and the graduate, the number of promoted students, supervision of scientific associations, initiatives in the field of cooperation with students and their organizations, as well as teaching facilities;

- organizational potential, including the ability to develop pro-efficiency and pro-quality intra-organisational projects, work for the university environment, and the organization of the internal units' operation.

**The development perspective (D):**

- acquiring and conducting development projects, fully informed at HR development, as well as management structures and infrastructure;
- development level and evolution of IT systems supporting the implementation of all university processes;
- level of organizational maturity.

The diversity of perspectives enables the development of multidimensional reports for managerial purposes. It also determines the architecture of the data aspect of the IMCSHE. The class diagram in Fig. 1. represents the simplified hierarchical structure of a university. Also, it shows how reports are generated. At this point, there were four types of reports distinguished concerning the identified perspectives, namely:

- Financial report,
- Customer report,
- Internal process report,
- Development report.

However, a new type of report could be defined if needed, since the generalization set is not complete.

The composition used between all levels of university management indicates that reports generated for each level of the university are used to calculate reports for higher university structures, e.g. in the case of a financial report, the sum of all expenses for projects in a given division will give the expense value in the financial report for this division. One of the project approach's main assumptions is that business as usual (BAU) is also treated as a separate project. At first, we distinguished two types of organization unit: Faculty and Administration. The lower level is a division, which serves as a universal type of second-level descendant in the structure. The division report is created as a result of the aggregation of projects, i.e. the lowest level of university structure decomposition. Reports are complex structures built from various fields, for each of which a separate aggregation function needs to be defined. At the top of the hierarchy is the university, which has reports derived from lower structure levels. The aggregation function allows us to derive the final report based on a chosen organizational structure level due to such a hierarchical structure. In particular, we do not need to include all levels in defining aggregation, instead only focusing on the minimum needed set of projects. However, each report is based on an arbitrary set of report types, i.e., Financial report, Customer report, Internal process report, and Development report. All of them inherit from a report, which means that they share some properties, but at the same time, they focus on different perspectives. Final Report should reflect all the mentioned perspectives and creates summarised (including different measurable results) view on the university state.

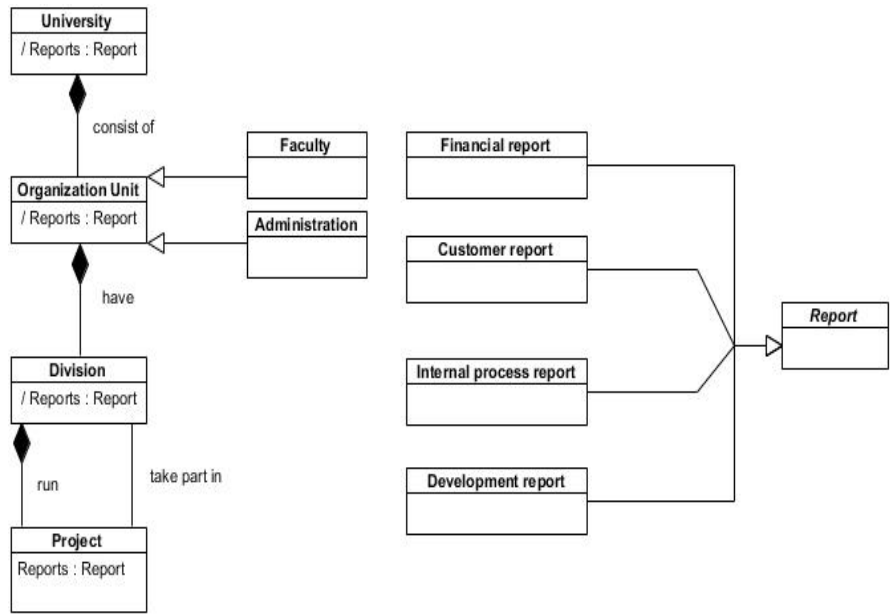

Fig. 1. The simplified hierarchical structure of a university

The component diagram (Fig. 2) shows the system's concept with a strong emphasis on data sources, i.e., the university's core IT systems. The integrated system communicates with each source component through a dedicated API and obtains the data from each of them. At first, the data feeds the data warehouse, where it is processed, and then the aggregated data is sent via a dedicated interface to the target iSIC component. Thus, the data warehouse is a crucial part of the system. The quality and structure of data delivered to the warehouse is undoubtedly the key indicator of system performance. In our concept, we focused on the following data sources:

- Employee evaluation system – provides data about employee results;
- USOS – provides data related to teaching, such as student surveys or data on financial settlements with students;
- Lecture planning system – a system for sharing data on the allocation of items and teachers to rooms;
- Publication repository – provides data on employees' publications;

- ERP – a system focusing on data such us: HR and payroll, organizational structure, and cost centres;
- Budget Planning System (SIC) – a system used to define unit budgets (tasks and project budgets);
- Expenses Control – a system focusing on tracking expenses;
- Project repository - shares information on scientific, R&D, and organizational projects;
- Infrastructure Management – provides information regarding the infrastructure.

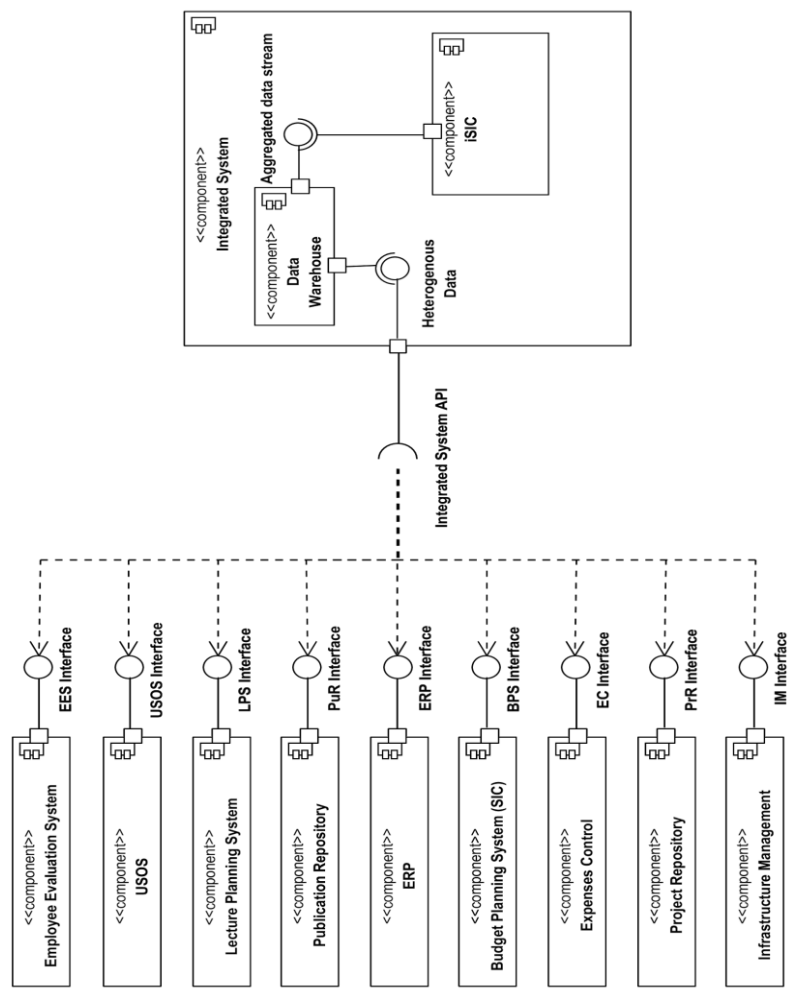

Fig. 2. A concept of the Management Control System for Higher Education.

Such diverse data sources allow generating distinguished types of reports by the iSIC component. The iSIC component consists of the artificial intelligence module. We use the Learning Intelligent Distribution Agent Architecture (LIDA), developed by Cognitive Computing Research Group (CCRG) [13, 14]. CCRG also developed the software framework of LIDA (in Java and Python). The reason for choosing this architecture is our many years' scientific experience in applying this architecture in management information systems [15, 16]. LIDA is a cognitive architecture, performing tasks within a cognitive cycle. The knowledge is represented by a "slipnet" - a semantic network with nodes' and links' activation level. The learning process is permanently performed. LIDA consists of the following modules:

- sensory memory,
- perceptual memory,
- episodic memory,
- declarative memory,
- workspace,
- attentional codelets,
- global workspace,
- action selection,
- sensory-motor memory.

The communication between modules is handled by codelets - mobile software programs processing information. LIDA's functioning was described in detail in [Franklin et al. 2016, McCall et al. 2020]. LIDA's advantage is related to the ability to process both symbolic and numerical knowledge (a slipnet allows the representation of such kinds of knowledge). This form of knowledge representation is beneficial in the areas of management and economics [17]. However, the disadvantage of LIDA is its complexity when processing large amounts of knowledge. Moreover, it entails high resource consumption. Therefore, we propose extending the LIDA architecture with a machine learning module, making it possible to process faster numerical knowledge (which cannot be represented as a slipnet, e.g. knowledge related to the prediction of costs, or the number of students). The extension is presented in Fig 3.

The machine learning module is connected to episodic memory (this memory stores data related to past events) and workspace memory. Input data for the machine learning module is required from the historical data and workspace memory. Machine learning module consists of 2 units:

- Experimental platform,
- Operational platform.

The experimental platform is the first unit, where ideas for Machine learning analyses are examined. Models are created, tested, validated and if necessary, tuned to archive better results. After a technical perspective of a model is verified, the business attractiveness of results is also evaluated. Positively verified models are documented, virtualized and deployed to the operational platform, where then return on a regular basis and generate results. The operational platform is also monitored, and it is checked if results accuracy does not drop below defined thresholds. If such a situation happens, the experimental platform is employed, and the model is again verified, tuned and in a corrected version again deployed to the operational platform.

The results are sent to the workspace memory. The results (e.g. predictions) are taken into consideration in forming a current situational model in the global workspace, which influences action selection and action forming.

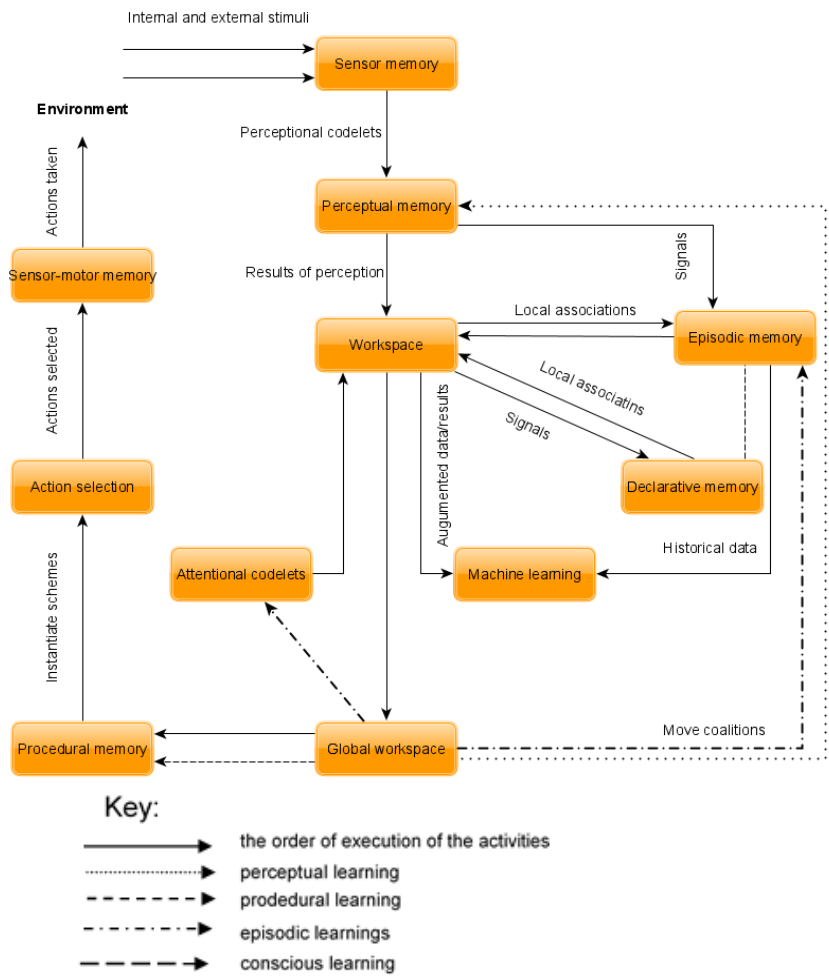

Fig 3. The extended architecture of LIDA.

Next, the set of alternative decisions is presented to the user. Each of decision has the level of confidence. Also consensus of the set of decisions is calculated [15].

## Conclusions

Universities try to develop following the strategies they adopted, which will enable long-term activities plans. The university management must constantly analyze the information coming from available databases pertaining to many areas and its surrounding environment. This article presents a concept of an intelligent management control system for higher education. The essence of this solution is a combination of an MCS's ideas, the balanced scorecard, and machine learning. The premise of this system is a controlling-oriented approach to supporting university management. The executed operational tasks must serve to implement the adopted strategy. The proposed system is expected to involve long-term measurement and management in key areas, i.e. to conduct research, providing education to students, and cooperating with the socio-economic environment.

Furthermore, it should support the university's management staff in carrying out various types of financial analyses so that the tertiary institution's activity does not bring financial losses. The university's MCS is based on the Balanced Scorecard, which comprises four perspectives: The Financial Perspective, the Customer Perspective, the Internal Process Perspective, and Development Perspective. The Balanced Scorecard is one of the methods used to support knowledge management, which is important because the university's employees' knowledge constitutes its essential capital. As far as universities are concerned, adaptation to the technological change, effective knowledge sharing and communication, operational efficiency, time saved due to knowledge sharing, and the growth and enhancement in the collaboration culture are becoming essential elements of knowledge management. Especially knowledge sharing allows generating new ideas, improving organizational performance, and helping employees to stay motivated. The Balanced Scorecard makes it possible to generate various reports for the university management with a focus on four perspectives.

The proposed concept of an intelligent control system requires an appropriate IT system architecture. A basis is a typical approach applied in Business Intelligence class IT systems, i.e. the transformation of data in various heterogeneous databases into information and knowledge for the management staff. There are many databases within the university (mentioned in the previous section) from which the necessary data will be obtained and loaded into a data warehouse after using ETL tools (initial concepts addressed to educational/customer area was proposed in [18]). Aggregate data located in a data warehouse is the primary resource for an Intelligent Management Control System for Higher Education. The use of the Learning Intelligent Distribution Agent Architecture and the machine learning module application are important elements of the proposed system. The use of these elements makes our proposal different from a typical Business Intelligence class system,

The proposed Intelligent Management Control System for Higher Education, applying four perspectives following the Balanced Scorecard and machine learning, allows the development of scenarios for future decisions that can be generated to ensure the adopted development strategy's execution. Scenarios may concern the processing of information from databases existing at the university. They may also include data and information from the university's environment, i.e. the demographic situation, changes in legal regulations, the demand for graduates with new qualifications, and competitive entities' activity.

**Acknowledgement.** The project is financed by the Ministry of Science and Higher Education in Poland under the programme "Regional Initiative of Excellence" 2019 - 2022 project number 015/RID/2018/19 total funding amount 10 721 040,00 PLN.